# Landau Levels of Topologically-Protected Surface States Probed by Dual-Gated Quantum Capacitance

Su Kong Chong[1], Ryuichi Tsuchikawa[1], Jared Harmer[1], Taylor D. Sparks[2] and Vikram V. Deshpande[1]*

[1]Department of Physics and Astronomy, University of Utah, Salt Lake City, Utah 84112 USA

[2]Department of Materials Science and Engineering, University of Utah, Salt Lake City, Utah 84112 USA

*Corresponding author: vdesh@physics.utah.edu

**ABSTRACT**

Spectroscopy of discrete Landau levels (LLs) in bulk-insulating three-dimensional topological insulators (3D TIs) in perpendicular magnetic field characterizes the Dirac nature of their surface states. Despite a number of studies demonstrating the quantum Hall effect (QHE) of topological surface states, quantitative evaluation of the LL energies, which serve as fundamental electronic quantities for study of the quantum states, is still limited. In this work, we explore the density of states of LLs by measuring quantum capacitance ($C_Q$) in a truly bulk insulating 3D TI *via* a van der Waals heterostructure configuration. By applying dual gate voltages, we access the individual surface states' LLs and extract their chemical potentials to quantify the LL spacings of each surface. We evaluate the LLs' energies at two distinguished QH states, namely dissipationless ($v= \pm1$) and dissipative ($v= 0$) states in the 3D TI.

The topologically-protected linearly dispersing surface states of three-dimensional topological insulators (3D TIs) offer a route for exploration of quantum phenomena such as the topological exciton superfluid, fractional charges and Majorana fermions in quantum regimes.[1-3] In a strong perpendicular magnetic field, the surface Dirac fermions localize in cyclotron orbits, forming discrete energy states known as Landau levels (LLs). This gives rise to the quantum Hall effect (QHE) in magnetotransport, where the Hall conductivity is quantized at integers as a result of half-integers from each (top and bottom) surface.[4-6] Alternatively, quantum capacitance ($C_Q$) can also be exploited to study the LLs electronically.[7-10] Different from quantum transport which is only sensitive to edge modes, $C_Q$ provides a probe of the charge states in the bulk of the surfaces.

$C_Q$ is an ideal measure of the thermodynamic density of states (DoS) for low carrier density and gate-tunable materials.[8-10] Nevertheless, it is relatively less explored in 3D TIs in the past as compared to quantum transport primarily due to the shortcoming of a proper 3D TI candidate. Previous studies of $C_Q$ in $Bi_2Se_3$ revealed the Dirac-like surface states[11] and Shubnikov-de Haas-like quantum oscillations in magnetic field.[12] However, $Bi_2Se_3$ is known to exhibit substantial carrier doping in bulk, and extra trapped states induced by intrinsic defects.[11,13] More recently, $C_Q$ in high mobility 3D TI based on strained HgTe was reported,[14] showing more intrinsic bulk band and clear LL quantization. Yet, the narrow bulk bandgap in strained HgTe (~20 meV), which is equivalent to lowest LL energy spacing at ~4T, limits the study of LLs only to low magnetic fields. Alternately, quaternary Bi-based TI compounds in the form $Bi_{2-x}Sb_xTe_{3-y}Se_y$ exhibit a relatively large bulk bandgap (~0.3 eV),[4,15] and fully suppressed bulk conduction at low temperatures, and serve as ideal candidates for probing the surface state LLs in $C_Q$.

In this work, we study the $C_Q$ of $BiSbTeSe_2$ (BSTS) devices based on van der Waals (vdW) heterostructures by assembly of graphite/hBN/BSTS/hBN/graphite sandwiched layers.[6] Such a configuration acts as three parallel plate capacitors in series, where chemical potentials of the topological surface states can be controlled by applying voltages through individual graphite (Gr) layers. Thin hBN and BSTS flakes are used in device fabrication to achieve large geometric and TI bulk capacitances, such that the measured signals will be dominated by $C_Q$. The devices are fabricated with Hall bar configured contact electrodes for both transport and capacitance measurements in a single device. Magnetic field up to 18 T is applied to the BSTS vdW devices to study the formation of LLs in topological surface states.

**RESULTS AND DISCUSSION**

Fig. 1(a) shows the optical image of a BSTS vdW five-layer heterostructure device for both quantum transport and capacitance measurements. The device is made of a thin BSTS flake (~17 nm), where the bulk capacitance is large enough yet the hybridization of the surfaces is insignificant. Low-temperature electronic transport of the BSTS device is measured through the Hall bar electrodes. Fig. 1(c) shows four probe resistance ($R_{xx}$) of the BSTS as a function of bottom-gate voltage ($V_{bg}$) measured at 0.3 K. The $R_{xx}$ peak at $V_{bg}$~-0.5V is assigned to charge neutrality point (CNP), where the peak position indicates a lightly *n*-doped BSTS. By applying dual gate voltages, the $R_{xx}$ is controlled to its maximum, corresponding to the overall CNP (interception of top and bottom CNPs) of both surfaces (inset in Fig. 1(c)). The diagonally skewed $R_{xx}$ maximum indicates strong capacitive coupling between the top and bottom surface states.[16,17]

Fig. 1(b) shows a schematic of the configuration for capacitance measurements. The top (bottom) surface of the BSTS forms a parallel plate capacitor with the top (bottom) Gr conducting layer, in which hBN acts as a dielectric layer. Symmetric capacitance ($C_S$) signals of the top and bottom surface states are measured by applying two in-phase AC excitation voltages to the top and bottom Gr layers, and detecting the unbalanced current ($\delta I$) through a capacitance bridge.[18] Fig. 1(d) presents the $C_S$ of the same BSTS device (Fig. 1(a)) as a function of $V_{bg}$, measured in a separate cooling cycle. The $C_S$ is tuned through a minimum which coincides with the $R_{xx}$ peak at the CNP. The minimum value of $C_S$ is associated with the low DoS near the CNP.[9,10,19] The $C_S$ tends to saturate at large $V_{bg}$ as the contribution of $C_Q$ to the total capacitance decreases at high charge densities. The dual-gated map of $C_S$ inserted in Fig. 1(d) shows the same diagonal feature as the $R_{xx}$ map. The diagonal feature splits into two corresponding to CNPs in each of the parallel surfaces as the charge density tuned away from the overall CNP.

We first inspect the transport of the BSTS device as functions of magnetic field and a single gate voltage, namely $V_{bg}$ as shown in Fig. 2. The vanishing longitudinal resistance ($R_{xx}$) and quantized Hall resistance ($R_{xy}$) $h/\nu e^2$ (with integer values of $\nu$) in magnetic field are a consequence of LL formation. The linear traces emanating from the CNP as shown in the $R_{xx}$ and $R_{xy}$ color maps signify a direct proportionality of magnetic field (B) *versus* bottom surface charge density ($n_b$). This is an indication of the Dirac dispersion of the LLs residing in topological surface states.[20,21] The slopes of the B *versus* $n_b$ are extracted from the $R_{xx}$ color map as $3.5\times10^{-11}/\nu$ T·cm² (hole) and $-3.3\times10^{-11}/\nu$ T·cm² (electron), which are close to theoretical relation of $\frac{dB}{dn} = \frac{h}{\nu e} \approx 4.1\times10^{-11}/\nu$ T·cm².[22,23] To further confirm the origin of the formed LLs, we plot the LL index (N) *versus* 1/B curves at different gate voltages ($V_{bg}-V_D$), and extract the y-intercept ($\beta$) as presented in Fig. S1. The converging curves are linearly fitted and yield $\beta$ of ~0.48±0.04, which provides strong evidence of the $\pi$ Berry phase arising in the Dirac surface states.[24,25]

Consistent with transport, the $C_S$ develops into symmetric dips coinciding with the minima and plateaus in $R_{xx}$ and $R_{xy}$, respectively, about the CNP. The sharp capacitance dips developing at high magnetic field are a strong indication of the diminishing DoS due to large cyclotron gaps between the discrete LLs.[9,10,19] To further study the evolution of capacitance in the magnetic field, we plot the magnitude of $R_{xy}$, $R_{xx}$, and $C_S$ as a function of the magnetic field for $v= +1$ and $-1$ LLs in Fig. S4. A noticeable change in $C_S$ prevails as the $R_{xy}$ approaches quantum limit of $\pm h/e^2$ and the $R_{xx}$ diminishes to zero. The magnitude of $C_S$ gradually deviates from a linear function with the further increase in magnetic field, denoting a nonlinear proportionality of quantum capacitance to magnetic field.

Color maps of magnetotransport and magnetocapacitance of the BSTS device as a function of dual gate voltages are analyzed in Fig. 3. The dual-gating provides independent control of chemical potentials of the top and bottom surfaces. By tuning the two surface states across CNPs, four-quadrant plateaus develop around the overall CNP. $R_{xx}$ (Figs. S7(a) and (b)) forms well-defined minima (maxima) in the parallel (counter)-propagating quadrants[6] separated by the clear onset of the LLs. The boundaries around the overall CNP are traced by dashed lines and labeled with the LL indices of top ($N_t$) and bottom ($N_b$) surfaces in longitudinal conductivity ($\sigma_{xx}$) maps in Fig. 3(a). Hall conductivity ($\sigma_{xy}$) (Fig. 3(b)) develops into rhombus-shaped QH plateaus at integer LL filling factors ($v$), corresponding to the sum of half-integers for each surface ($v= v_t+v_b = (N_t+½)+(N_b+½)$),[4,6,26] in the $\sigma_{xx}$ minima regions. The four quadrants are assigned to two symmetric $v= 0$ (white); and antisymmetric $v= -1$ (blue) & $+1$ (red) QH plateaus. The $\sigma_{xy}$ as a function of $V_{bg}$ swept through different $V_{tg}$ values reveals a set of quantized plateaus with a step height of $e^2/h$ as indexed in line profiles in Fig. 3(b).

The dual-gated magnetocapacitance grants access to LLs in the bulk of the parallel paired topological surface states. The dual-gated $C_S$ reveals well-formed capacitance dips about the overall CNP as shown in Fig. 3(c). The rhombus-shaped capacitance dips coincide with the four quadrants QH plateaus in $\sigma_{xy}$. The most prominent dips between lowest LLs ($N_b$, $N_t$= +1, 0 and -1) are clearly resolved in the color map. The line profiles of $C_S$ as a function of $V_{bg}$ cutting through the three different $V_{tg}$ show nearly identical capacitance dips emerging for the $\nu= 0$ and $\pm 1$ plateaus, suggesting a similar cyclotron gap size formed in these states. Similar behavior is also observed in dual-gated magnetocapacitance of additional BSTS devices (Figs. S5 and S6). An effective way to confirm the cyclotron gap of LL-induced $C_Q$ dips is by measuring the out-of-phase current signals, known as dissipation. Symmetric dissipation ($D_S$) depicted in Figs. S7(c) and (d) detects pronounced dissipative signals at the onset of LLs, and the signals vanish inside the LL gaps. Such $D_S$ features were also observed in graphene LLs.[18] The relatively large background dissipative signals in capacitance are attributed to the resistive nature of the BSTS surface states in the entire density region (refer to zero field transport data). By flipping the sign of the AC excitation voltage applied to the top Gr layer, dual-gated anti-symmetric capacitance ($C_A$) were measured (Fig. 3(d)). The opposite phase AC excitation voltage is adjusted to nullify the signal at high charge densities. As $C_A$ probes the antisymmetric combination of the top and bottom surface capacitance, a color map of $C_A$ (*versus* dual gate voltages) shows opposite signs for the background signal across the dashed line (zero electrical polarization). The $C_A$ color map reveals an anti-symmetric feature between $\nu= 0$ (red) and $\nu= \pm 1$ (blue) QH plateau states, which accounts for the difference between the quantum capacitances of the respective LLs.

The total capacitance (C) can be expressed as $1/C = 1/C_g + 1/C_B + 1/C_Q$, where $C_g$, $C_B$, and $C_Q$ are geometric, BSTS bulk and quantum capacitances, respectively.[12,27,28] Note that we minimize parasitic capacitances from the measurement lines through the use of well-shielded coaxial cables, local Gr gates, and highly-resistive intrinsic Si substrates. The $C_B$ strongly depends on the thickness of the BSTS, which can be estimated from dual-gated $V_{tg}$ *versus* $V_{bg}$ plots using a capacitor charging model[15] (Fig. S8). The effective surface area of the BSTS flake of about 594 µm$^2$ (estimated from the optical image) yields a $C_B$ of ~23 pF, about two orders of magnitude larger than the measured C. Thus, we can disregard the contribution of $C_B$ in the total capacitance. The measured C in the form of $C_S$ and $C_A$ is equivalent to the sum and difference between capacitances of top ($C_t$) and bottom ($C_b$) surfaces, respectively. The $C_t$ and $C_b$ are solved algebraically from the $C_S$ and $C_A$ to further study the individual surface states. Subsequently, quantum capacitance of top ($C_{Qt}$) and bottom ($C_{Qb}$) surface states can be extracted from the $C_b$ and $C_t$ using the relations as $1/C_{b(t)} = 1/C_{gb(gt)} + 1/C_{Qb(Qt)}$, where the geometric capacitance ($C_{gb(gt)}$) is calculated from the parallel plate capacitor relation, $C_g = \varepsilon_{hBN}\varepsilon_o A/d_{hBN}$ with dielectric constant of hBN taken to be 3. The $C_{Qb}$ and $C_{Qt}$ dual-gated maps are plotted in Fig. S9.

$C_Q$ is related to electronic compressibility as $C_Q = Ae^2 \frac{dn}{d\mu}$,[7-10] where A is the effective surface area, n and µ are the surface charge density and chemical potential, respectively. µ(n) can be obtained by integrating the electronic compressibility (dµ/dn) with respect to n.[9,10] In our calculations, we applied the charge density relation of bottom surface ($n_b$) based on the derivations in literature[15] as $n_b = C_{bg}\Delta V_{bg} + \left(1 + \frac{C_{bg}}{C_B}\right)C_{tg}\Delta V_{tg}$, where the first term is the density controlled by backgate and the second term encounters the capacitive couplings to the top-gate and top surfaces. The same derivation is used for the charge density relation of top surface ($n_t$) by switching the back- and top-gating terms. The chemical potential ($\mu_b$) as a function of charge density ($n_b$) of the bottom surface at different magnetic fields from 12 to 18 T are plotted in Fig. S10(a). The jumps in $\mu_b$ around the integer $N_b$ indicate the corresponding gaps between $N_b$= +1, 0 and -1 LLs developed in the bottom surface state. The LL energies of bottom surface state ($E_b$) for LL indices of $N_b$= +1 and -1 as a function of magnetic field are extracted from the step heights and plotted in Fig. S10(b). We note that $E_b$ for B< 12T is omitted as the capacitance dips are not well-resolved due to overlap of the dips with the minimum DoS at CNP. The transport data ($\sigma_{xy}$) as a function of magnetic field for the corresponding $\nu$= +1 and -1 plateaus are included in Fig. S3(a). $E_b$ values approach the theoretical model of LL energy of topological surface states $E_N=\text{sgn}(N)v_F\sqrt{2e\hbar|N|B}$ [29,5] at high magnetic field as the $\sigma_{xy}$ reaches the complete development of ±1 $e^2$/h. The deviation of $E_b$ from the Dirac LL energy equation at low magnetic field is attributed to the disorder broadening in the LL bands. To analyze this effect, we extracted the disorder broadening parameter, $\Gamma_{SdH}= \hbar/\tau_q$ [30] from the quantum lifetime ($\tau_q$) based on temperature and magnetic field dependence Shubnikov-de Haas (SdH) oscillation amplitude analyses as presented in Fig. S2. The obtained $\Gamma_{SdH}$ of ~7.1 meV agrees with the LL broadening at low magnetic field. We note that the $N_b$= ±1 LL chemical potentials steps of ~40-42 meV are quantitatively more accurate compared to the thermal activation energy (~6 meV) of the same LL states at 31 T as reported by Xu *et al.*,[4] and comparable to the localized DoS probed by scanning tunneling microscopy.[22] Also, the LL energy estimation varies with the excitation frequency (Fig. S12(a)) and is affected by the parasitic capacitance in the measuring lines (Fig. S12(b)). Unless specified, the following analyses are based on measurements at 69.97 kHz which is in range limit of the bandwidth of the lock-in amplifier.

Next, we turn to discuss the LL-modified chemical potentials of two (top and bottom) surface states at constant magnetic field. Fig. 4(a) presents a color map of $\mu_b$ as a function of top and bottom charge densities at 18 T. The dual-density $\mu_b$ map is generated by integrating the reciprocal of $C_{Qb}$ over $n_b$ for the entire $n_t$ range, with zero $\mu_b$ (white area) manually fixed at $N_b= 0$ LL ($n_b \approx 0$ /cm$^2$) as drawn in the figure. We focus on the $\mu_b$ in charge density regions corresponding to the four quadrant QH plateaus around the overall CNP. The $\mu_b$ *versus* $n_b$ data for LL filling factors $v=$ (*i*) $0^{\pm}$, (*ii*) +1, (*iii*) -1, and (*iv*) $0^{\mp}$ (as labeled in the color map) are plotted in Fig. 4(b). The notations of $0^{\pm}$ and $0^{\mp}$ are used for ($v_t$, $v_b$) of (+½,-½) and (-½,+½), respectively, to distinguish the two $v= 0$ states. The comparable gaps for the $0^{\pm}$ ($0^{\mp}$) and -1 (+1) states indicate that the chemical potential of bottom surface lie in the same LL *i.e.* $N_b=$ -1 (+1). Similar analyses are performed on $C_{Qt}$ to extract the $\mu_t$ *versus* $n_t$ relations at different $n_b$ (Fig. S11). By separating $E_b$ and $E_t$ from our dual-gated capacitances, we obtain the LL energies of individual surfaces at different QH states. The results of $E_b$ (black dots) and $E_t$ (red dots) for the corresponding $N_b$ and $N_t$ are plotted in Fig. 4(c). In $v=$ +1 (-1) states, the $E_b$ and $E_t$ display the same sign for both $N_b$ and $N_t$ because the chemical potentials of both surfaces reside in electron (hole) LLs. This is consistent with the origin of parallel-propagating states in conduction. Conversely, the $E_b$ and $E_t$ reveal nearly equal magnitude and opposite sign for the $v= 0^{\pm}$ ($0^{\mp}$) states due to the opposite occupation of $N_b=$ -1 (+1) and $N_t =$ +1 (-1). The LL energies from the top and bottom surfaces balance out and give rise to the $v= 0$ states.

In 3D TIs, the $v= 0$ quantum states are a consequence of two parallel surfaces occupying opposite LL fillings, corresponding to a counter-propagating state in conduction. Magnetotransport shows that these are dissipative quantum states as indicated by finite $R_{xx}$ (Fig. S7(b)), and residual $\sigma_{xx}$ (Fig. 3(a)) in the states. The $C_A$ map (Fig. 3(d)) captures the difference in quantum capacitance (or DoS) between the dissipative $v= 0$ states and dissipationless $v= \pm 1$ QH states. To further analyze the difference between these states, we plot the $E_t$ as a function of $\sqrt{B}$ for different $v$ states in Fig. 4(d). The $E_t$ is linearly fitted to the LL energy relation, which gives the Fermi velocity ($v_F$) of ~3.2-3.4×10$^5$ m/s for $v=\pm 1$, and ~2.6-2.9×10$^5$ m/s for $v=0$ states. The small difference in $v_F$ for the equivalent QH states could be related to the counterpropagating nature of the $v=0$ states which enhances scattering between surfaces[31] and reduces $v_F$ in the states. These $v_F$ of BSTS are consistent with similar Bi-based 3D TI compounds,[4,23,32,33] despite the magnitude being smaller than $v_F$ obtained from angle-resolved photoemission spectroscopy.[34] In addition, $E_t$ as a function of temperature measured at 18 T for the different $v$ are presented in Fig. 4(d). $E_t$ for $v=0$ and $\pm 1$ states decrease with increase in temperature and eventually reach the same energy level beyond 10 K, indicating a slightly different thermal activation between these states. The temperature dependence of $E_t$ is significantly different from the Hall conductivity of BSTS where the quantization is preserved up to 50 K as shown in Fig. S3(b) in consistent with literature.[4] The stability of QH states at higher temperature suggests that the LL bands remain robust up to 50 K. Therefore, the large response of $E_t$ with temperature in the lowest LL states could arise from a different origin such as mid-gap impurity states in TI surfaces[35] or possibly developing interacting states which are correlated with our smaller $v_F$.[36] This suggests that capacitance is more sensitive to the gap states in 3D TI compared to transport. Higher quality BSTS devices with significant reduction in disorder are required to resolve the extra features in these LLs.

**CONCLUSIONS**

In summary, we studied the quantum capacitance of topological surface states in a strongly bulk-insulating 3D TI in a dual-gated configuration. The Dirac nature of surface state LLs is confirmed by their linear B *versus* n relation, and β~0.48 corresponding to π Berry phase in LL fan diagrams. By integrating the inverse DoS over a charge density range, we obtained the energy spacings of the lowest LLs corresponding to the fully-quantized QH states. The largest cyclotron gap sizes of ~44-46 meV at 18 T agree with the theoretical relation of Dirac LL energies for N= ±1 LLs. The dual-gated capacitance with independent excitation source applied separately to two surfaces allows the individual probing of surface states. The opposite LL energies for top and bottom surfaces give rise to the $v=0$ quantum states. We observed lower Fermi velocity and thermal activation in $v=0$ states as compared to the conventional $v=\pm 1$ states. Capacitance measurement is promising for further study of the gapped quantum states such as the excitonic superfluid and fractional quantum Hall states existing in LL regimes.

## METHODS

**Device fabrication.** The vdWs heterostructures of BSTS/hBN/Gr devices were fabricated by a dry transfer method using a micromanipulator transfer stage.[6] An undoped Si wafer coated with 300-400 nm thermally oxidized layer was used as a substrate to avoid parasitic capacitance from the substrate. Bottom hBN/Gr layers were transferred onto the substrate, followed by annealing in Ar/$H_2$ environment at 360-380°C for 2 hours to provide a clean base for BSTS. BSTS thin flakes were exfoliated from a bulk crystal grown by a vertical Bridgman furnace,[26] and transferred onto the hBN/Gr using the transfer stage. The contact electrodes were made using a standard electron beam lithography process, followed by Cr/Au (2 nm/25 nm) deposition in an electron beam evaporator. The dual-gated device process was completed by stamping of hBN and Gr flakes onto the leads patterned BSTS.

**Transport measurement.** Low-temperature transport measurements were performed in a variable temperature helium 3 insert (base temperature of 0.3 Kelvin) equipped with a superconducting magnet up to 18 Tesla, based at National High Magnetic Field Laboratory. Two synchronized Stanford Research SR830 lock-in amplifiers were used to measure the longitudinal and Hall resistances concurrently on BSTS devices in Hall bar configuration. The lock-in amplifiers were operated at a frequency of 17.777 Hz. The device was typically sourced with a constant AC excitation current of 10-50 nA. Two Keithley 2400 source meters were utilized to source DC gate voltages separately to the top and bottom Gr gate electrodes.

**Capacitance measurement.** Capacitance was measured using a capacitance bridge configuration connected between the BSTS device and the parallel gold strip as a reference capacitor. Two synchronized (at a frequency of ~10-70 kHz) and nearly equal-amplitude AC excitation voltages (range of 10-40 mV) were applied separately to the top and bottom Gr gates, whose relative magnitude was chosen to match the ratio of geometric capacitances of top and bottom surfaces. The AC excitations are applied in phase for $C_S$ and out of phase for $C_A$ measurements. A third AC excitation voltage was applied to the reference capacitor with the amplitude set to null the measured signal at high gate voltages where the signal is saturated with the gate voltages. The reference capacitance ($C_{ref}$) was calibrated to be 267 fF using a standard capacitor (Johanson Technology R14S, 1 pF). The $C_S$ and $C_A$ data were acquired by monitoring the off-balance current ($\delta I$) at the balance point as the DC gate voltages were changed. The unbalanced current ($\delta I$) is converted to the capacitance difference ($\delta C$) using a standard relation $\delta C = \delta I / \omega V_{ex}$, where $\omega = 2\pi f$ and $V_{ex}$ are the applied frequency and excitation voltage, respectively. The $\delta C$ is then added to the $C_S$ or $C_A$ obtained from the capacitance bridge at the null point. The $C_S$ and $C_A$ are related to the capacitances of top ($C_t$) and bottom ($C_b$) surfaces as $C_S = C_b + C_t$ and $C_A = C_b - C_t$.[18]

**ASSOCIATED CONTENT**

**Supporting Information**. This material is available free of charge *via* the Internet at http://pubs.acs.org. Further information on the magnetic field and temperature dependence transport properties (Figures S1-S4), dual-gated magnetotransport and magnetocapacitance (Figures S5-S9), chemical potentials and Landau level energies of topological surface states (Figures S10-S11), and the effects of excitation frequency and parasitic capacitance on Landau levels' energy analysis (Figure S12).

## AUTHOR INFORMATION

**Corresponding Author:**

*E-mail: vdesh@physics.utah.edu

**Author Contributions**

S.K.C. and V.V.D. designed, conducted the experiments and prepared the manuscript. J.H. involved in the device fabrication. R.T. helped in the setup of the capacitance measurements. T.D.S. provided single crystal $BiSbTeSe_2$ topological insulator. All authors contributed to the discussion of results and approved the final version of the manuscript.

## ACKNOWLEDGMENTS

The authors thank Haoxin Zhou for valuable discussions. This material is based upon work supported by the National Science Foundation the Quantum Leap Big Idea under Grant No. 1936383. A portion of this work was performed at the National High Magnetic Field Laboratory, which is supported by National Science Foundation Cooperative Agreement No. DMR-1644779 and the State of Florida.

**FIGURES**

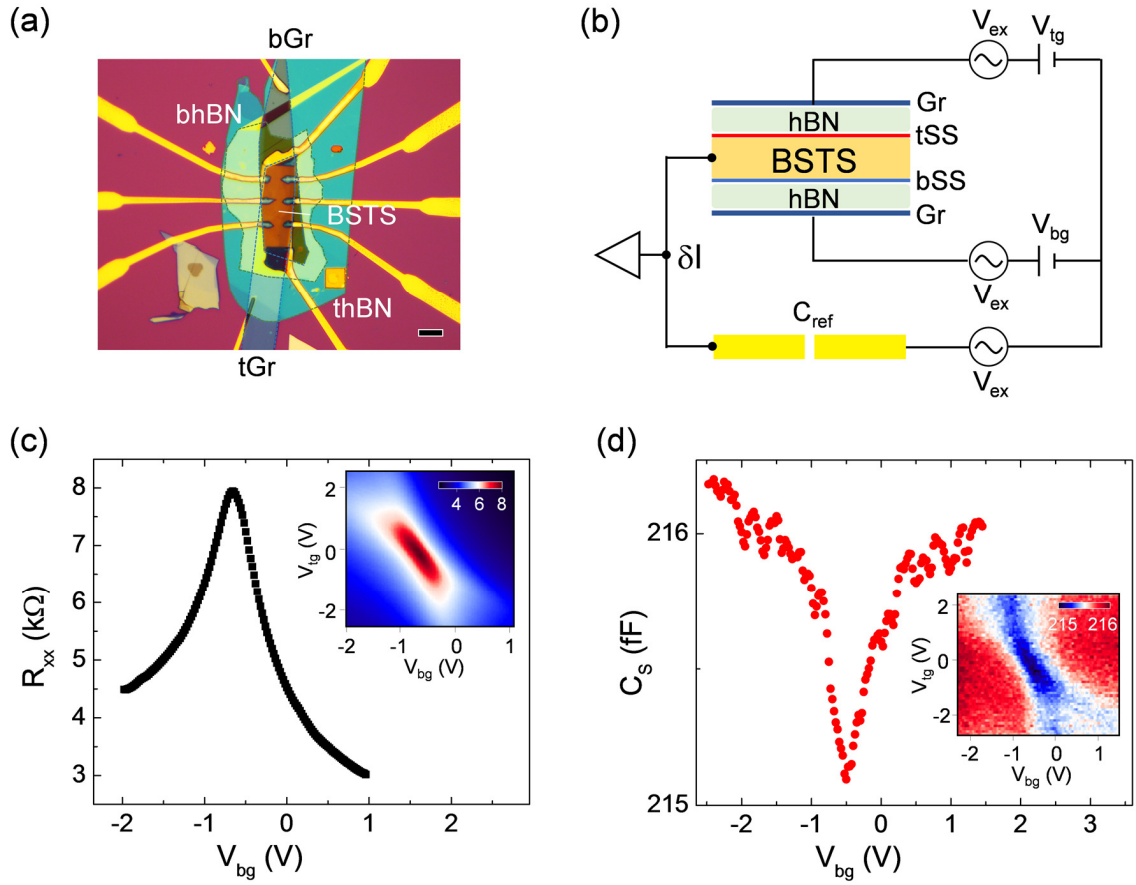

**Figure 1** (**a**) Optical image of a vdW heterostructure of BSTS/hBN/Gr device fabricated on an undoped Si substrate. The scale bar is 10 μm. (**b**) Schematic diagram of the dual-gated quantum capacitance measurement. (**c**) Four probe resistance ($R_{xx}$), and (**d**) symmetric capacitance ($C_S$) as a function of $V_{bg}$ for the BSTS device measured at temperature and magnetic field of 0.3 K and 0 T, respectively. Insets in (c) and (d) are the 2D color maps of the $R_{xx}$ and $C_S$, respectively, as a function of dual gate voltages. Color scales in insets of (c) and (d) are $R_{xx}$ and $C_S$ in units of kΩ and fF, respectively.

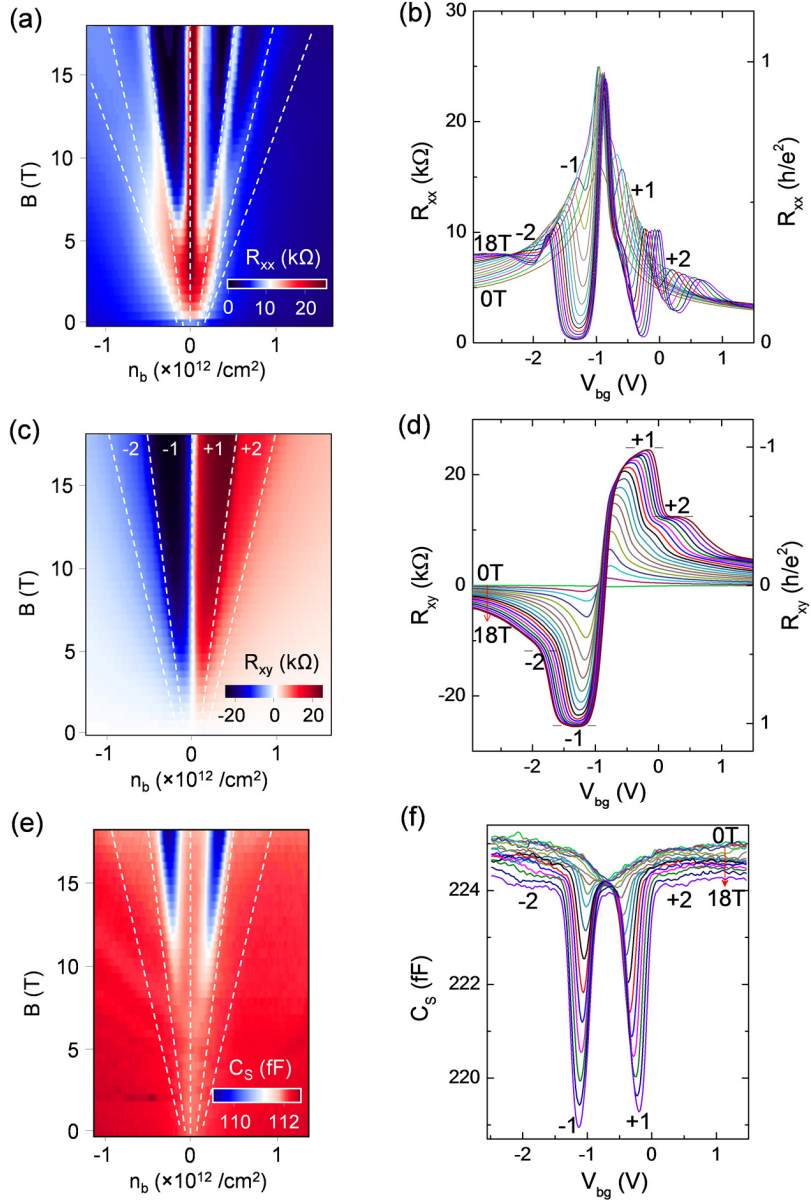

**Figure 2** Color maps of (**a**) $R_{xx}$, (**c**) $R_{xy}$, and (**e**) $C_S$ of the BSTS vdW heterostructure device as a function of magnetic field (B) and charge density ($n_b$) measured at 0.3 K. The $n_b$ is converted from gate voltage ($V_{bg}$) using the relation as $n_b = C_g(V_{bg} - V_D)$, where the $C_g$ and $V_D$ are gate capacitance (geometric capacitance of hBN dielectric and Gr layers) and peak position of the Dirac point, respectively. The QH plateaus and capacitance dips are symmetrical about the overall CNP ($n_b = 0$). The white dashed lines in (a), (c) and (e) trace the onset of the LLs about the Dirac point. Plots of (**b**) $R_{xx}$, (**d**) $R_{xy}$, and (**f**) $C_S$ curves as a function of $V_{bg}$ at different magnetic fields. The respective LL filling factors ($\nu$) are labeled in the $R_{xy}$ map (c), and the plots in (b), (d) and (f).

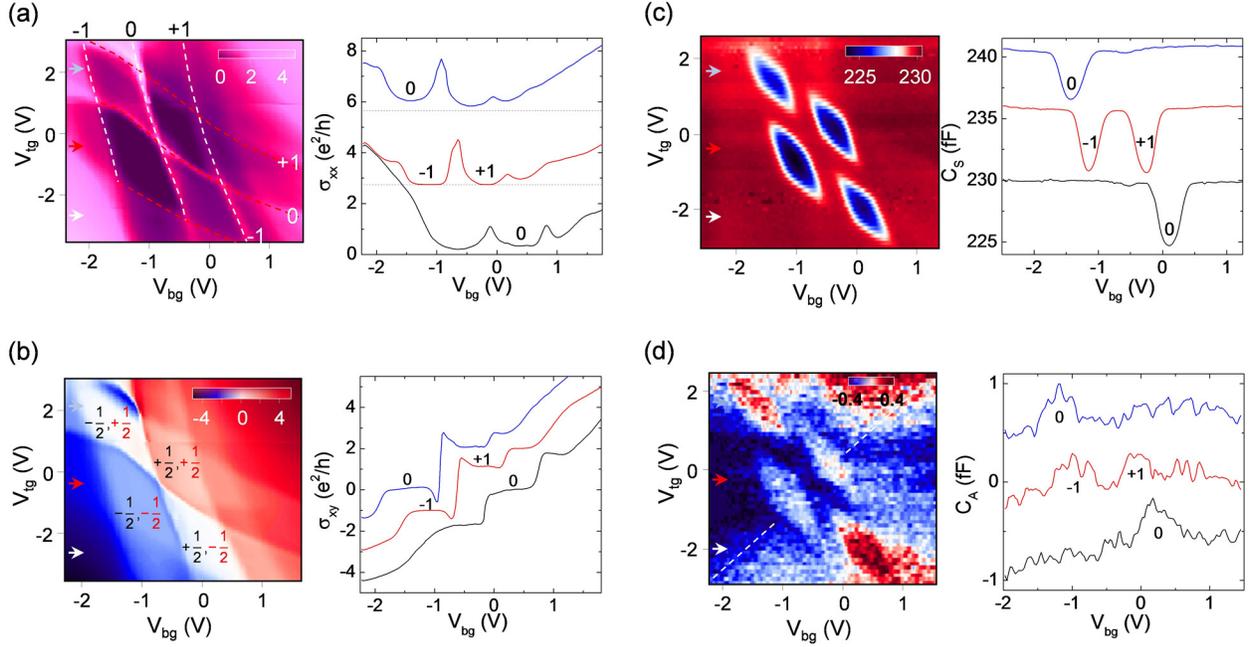

**Figure 3** Left: 2D color maps of (**a**) $\sigma_{xx}$, (**b**) $\sigma_{xy}$, (**c**) $C_S$, and (**d**) $C_A$ of the BSTS device as a function of dual gate voltages measured at temperature and magnetic field of 0.3 K and 18 T, respectively. Right: Line profiles of (**a**) $\sigma_{xx}$, (**b**) $\sigma_{xy}$, (**c**) $C_S$, and (**d**) $C_A$ as a function of $V_{bg}$ extracted at different $V_{tg}$ from the color maps. Labels and tracelines in color map (a) represent the LL indices of top (black) and bottom (red) surfaces. Dashed lines in $\sigma_{xx}$ plots in (a) indicate zeros on the y-axis in each plot. Indices in color map (b) are the top (red) and bottom (black) LL filling factors ($\nu_t$, $\nu_b$). Color scales in (a) & (b), and (c) & (d) are plotted in units of $e^2/h$ and fF, respectively. The line profiles in (c) and (d) are manually up-shifted for clarification.

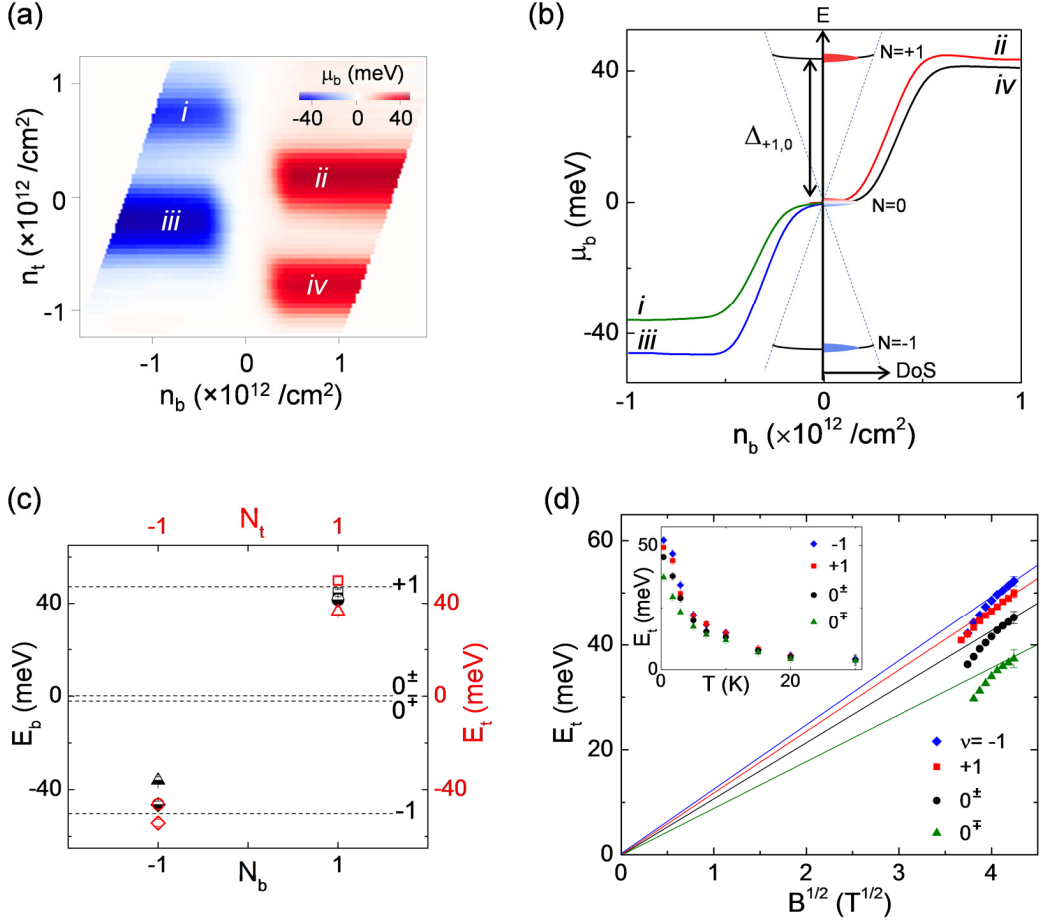

**Figure 4** (**a**) 2D color map of chemical potential of the bottom surface state ($\mu_b$) as a function of charge density of top ($n_t$) and bottom ($n_b$) surfaces. (**b**) Line profiles of $\mu_b$ as a function of $n_b$ at different $n_t$, as labeled in (a) color map. Inset in (b) is a schematic of the energy (E) *versus* density of states (DoS) diagram for the respective LLs of bottom surface. The LL energy spacing, $\Delta_{+1,0}=E_{+1}-E_0$, is labeled in the figure. (**c**) Plots of $E_b$ and $E_t$ *versus* $N_b$ and $N_t$ for $\nu=+1$ (square), -1 (rhombus), $0^{\pm}$ (triangle), and $0^{\mp}$ (circle). Dashed lines in (c) represent the average energy of $E_b$ and $E_t$ for different $\nu$. (**d**) Plots of $E_t$ as a function $B^{1/2}$ for different LL filling factors. The color lines are guide to the eye of the linear fit to LL energy relation. Inset in (d) is the $E_t$ as a function of temperature.

**TABLE OF CONTENTS GRAPHIC**

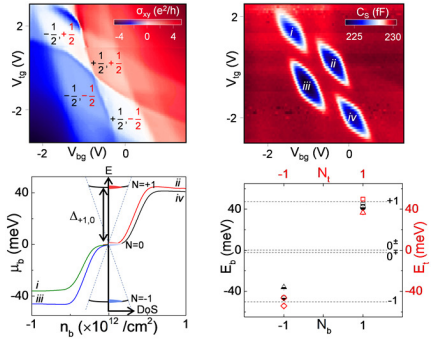